\documentclass[pre,twocolumn,notitlepage,amsmath,amssymb]{revtex4-1}

\usepackage{graphicx}
\usepackage{float}
\usepackage{placeins}
\usepackage[bf]{subfigure}

\begin{document}

\title{Can Spatiality Promote Diversity?}

\author{Bruno Messias$^1$}
\author{Filipi N. Silva$^1$}
\author{Cesar H. Comin$^2$}
\author{Luciano da F. Costa$^1$}
\affiliation{$^1$S\~ao Carlos Institute of Physics, University of S\~ao Paulo, PO Box 369, 13560-970, S\~ao Carlos, SP, Brazil\\}
\affiliation{$^2$Department of Computer Science, Federal University of S\~ao Carlos, S\~ao Carlos, SP, Brazil}


\begin{abstract}
Real-world dynamics running on networks can be characterized in terms of their respective diversity, or heterogeneity of state values.  Spatial networks can be understood as networks exhibiting limited small world characteristics.  In the present work we argue that network spatiality can enhance the diversity of respectively unfolding dynamics.  This also means that the small world property tends to reduce diversity. We illustrate this conjecture by simulating one type of Sznajd dynamics at the transient regime on Watts-Strogatz networks with varying rewiring levels.  The obtained results show a marked reduction of state diversity as spatiality is replaced by the small world property.
\end{abstract}

\maketitle

\section{Introduction}
\label{sec:introduction}
Modeling and studying real-world complex systems corresponds to one of the 
central research subjects currently, which is accounted by the fact that
most remaining scientific and technological problems have a predominantly
complex nature (e.g.~\cite{trabesinger_complexity_2011,kivelson_understanding_2018}).  An issue of particular 
interest concerns how the topology of interconnections in such systems may 
influence or constrain different types of dynamics unfolding on the network 
nodes.  For instance, we can model the world airport system as a complex network 
and understand the number of passengers at each airport at a given time $t$
as corresponding to the
dynamical state $s_i(t)$ of each respective node $i$.  Then, it would be 
interesting to try relating $s_i(t)$ to measurements of the network topology
such as the degree or betweenness centrality of nodes (e.g.~\cite{guimera2004modeling}).
This type of investigation relating the topology and dynamics of complex
systems has become one of the key issues in complex networks research,
being applied to many distinct problems such as epidemic spreading~\cite{pastor2015epidemic,vespignani2008dynamical}, synchronization~\cite{arenas2008synchronization}, neuronal dinamics~\cite{roxin2004self,kaiser2010optimal,comin2013shape}, network exploration~\cite{adamic2001search,sinatra2011maximal}, routing of information~\cite{yan2006efficient}, 
to name but a few applications~\cite{newman2010networks,costa2011analyzing,dorogovtsev2008critical}. This kind of analysis allows 
dynamical properties of a given system to be \emph{predicted} to some extent 
from the topological features of the respective interconnections, and vice versa. 

A natural approach to relate the topology and dynamics in a complex system
involves searching for relationships between measurements of these two
aspects.  For instance, one may be interested in finding how the
degree of a network relates to the respective state of the nodes
during a random walk dynamics (e.g.~\cite{lovasz1993random,da2007correlations,comin2014random}).  While it is known~\cite{lovasz1993random}
that, in non-directed networks, the activation of the nodes is fully correlated 
with the node degree, this correlation is typically lost in directed
networks, with a few exceptions~\cite{da2007correlations,comin2014random}.

In several real-world systems, there
is a particularly interesting dynamical property that has received relatively
less attention, namely the \emph{diversity}, \emph{variety} or 
\emph{heterogeneity} exhibited by the dynamical states associated to each of 
the network nodes.   For instance, we could be interested in studying the
variety of opinions of a given social group, e.g.~by taking the entropy
of these states, and then trying to relate this measurement with some topological
properties of the respective network.  This problem is particularly interesting
because it underlies several important properties of real-world systems such as
social networks (e.g.~memes~\cite{ramos2015does}), ecology (e.g.~number of species~\cite{dunne2002network}), biology (e.g.~functional diversity in the brain~\cite{senden2014rich}), economy (e.g.~number of competing companies~\cite{ferrary2009role}), fake news (e.g.~\cite{qiu2017limited}),
transportation (e.g.~passenger flow), among many other important systems.

The current work addresses the relationship between diversity and topology
from the perspective of taking into account the \emph{spatiality}
of the network connections as a particularly important parameter influencing the
diversity of the network dynamical states.  In particular, we quantify the 
latter dynamical
variable in terms of its entropy, while expressing the spatiality of the
respective network in terms of statistical properties (e.g.~average and standard
deviation) of its shortest paths.
We believe the above stated specific type of relationship between topology
and dynamics is particularly important because spatiality, by implying longer
shortest paths, acts as a kind of constraint to dissemination along time
and space of information throughout the network.  For example, the diffusion of 
opinions in the pre-internet era was much slower than nowadays because information
was spread by using less interconnected communication systems.  Because of
this property, it could be expected that most systems presenting stronger spatiality
would tend to exhibit a more diversified distribution of dynamical states at its nodes
for at least two possible reasons: (i) the transient would take 
longer~\cite{lang2018analytic,grabow2010small}; and (ii) less effective communication makes people less likely to be exposed to new ideas and eventually change their opinions.

In particular, we study these issues by considering the Watts-Strogatz network model,
as this allows a continuous gradation of spatiality from highly spatial 
structures as the lattice up to uniformly random networks exhibiting scant
spatiality.  We also adopt the Sznajd model~\cite{araujo2015mean} as a representative approach to opinion formation.  The reasons for this choice include the fact that this model is relatively simple, allows analytical treatment, reflects the influence of neighbors on the current opinion at each node, and allows good adherence to real data during the transient dynamics.  Nevertheless, we postulate that many other types of dynamics will also be characterized by the increase of diversity with spatiality.

The article is organized as follows. In Section~\ref{sec:spatiality} we describe the relationship between spatiality and the average shortest path length of networks, as well as the Watts-Strogatz model, which can be used to systematically generate networks having distinct degrees of spatiality. In Section~\ref{sec:dynamics} we present the Sznajd dynamics. Section~\ref{sec:results} contains the results of the study and the conclusions are presented in Section~\ref{sec:conclusions}.

\section{Spatiality and Shortest Paths}
\label{sec:spatiality}

Many real-world networks have the so-called \emph{small-world} property, meaning that the typical shortest path length between any two nodes grows proportionally to the logarithm of the number of nodes in the network~\cite{newman2010networks}. Here, an important distinction needs to be made regarding the shortest path length and the spatial (or metric) distance between two nodes. For unweighted networks, the calculation of the shortest path length does not depend on the spatial distance between the nodes. Nevertheless, in many real-world networks these two properties tend to influence each other. Such interplay between spatiality and topology is analyzed using spatial network models~\cite{barthelemy2011spatial}. 

In cases where the spatial position of the nodes bear strong influence on the connectivity, a node connects predominantly to other nearby nodes. We use the term \emph{high spatiality} henceforth to refer to such cases. For networks having high spatiality, nodes have strong restrictions on the neighbors that they can connect to. We aim at investigating if such restrictions can influence diversity in a dynamics taking place on the network.  In particular, we are interested in the hypothesis of diversity being proportional to spatiality.

The Watts-Strogatz (WS) network model~\cite{watts1998collective} is a simple model capable of generating networks having a specific degree of spatiality. The procedure for generating a WS network starts with a regular ring lattice, in which each node has degree $k$. Then, each edge of the network is rewired with probability $\rho$. If an edge is to be rewired, each endpoint of the edge becomes associated to a new node that is randomly drawn with uniform probability, thus creating a shortcut in the network. Notice that, while the edges that have not been rewired can be said to have short spatial length, the shortcuts are not influenced by the spatial positions of the nodes. Thus, for $0<\rho<1$ the network is composed of a mixture of spatial and non-spatial edges. For small $\rho$, a WS network has narrow degree distribution, large average shortest path length and large clustering coefficient. For $\rho=1$, an Erd\H{o}s-Rényi network is generated, having small average shortest path length and null clustering coefficient. Here we consider a slight variation of the WS model that starts with a lattice graph.

\section{The Sznajd Dynamics}
\label{sec:dynamics}

The Sznajd dynamics is a simple opinion propagation model that has been used to describe many real-word situations~\cite{castellano2009statistical}. Distinct versions of this dynamics have been defined in the literature. Here we adopt the complex network Sznajd model from~\cite{bernardes2002election}, which is defined as follows. Each node $u$ in the network can have an opinion $\sigma(t, u) \in \{0, 1,...N_o\} $ at time $t$, where $0$ indicates that the node is undecided. At each time step ($\text{MCT}$), a node $u$ is randomly selected with uniform probability. If node $u$ is undecided, nothing happens. Otherwise, one of its neighbors, $q$, is randomly selected with uniform probability. The following rules are then applied:

\begin{enumerate}
\item If node $q$ is undecided, then with probability $1/k_u$ it adopts the opinion of node $u$.

\item If nodes $u$ and $q$ have the same opinion, each neighbor of $u$ will adopt the same opinion with probability $1/k_u$. The same happens for neighbors of $q$, but with probability $1/k_q$.

\item If $u$ and $q$ have different opinions, nothing happens.
\end{enumerate}

In the following, we consider $N_o=2$.

\section{Results}
\label{sec:results}

In order to provide an illustration of the hypothesis that spatiality can enhance diversity along time, we performed the Sznajd dynamics described in Section~\ref{sec:dynamics} into Watts-Strogatz networks with varying rewiring probabilities $\rho$.  The networks have 1089 nodes and a total of 4000 realizations were performed for each configuration ($\rho$). The diversity was quantified in terms of the exponential of the entropy of the opinions at a given time MCT. Since the asymptotic equilibrium state for all rewiring values is one, we focus attention on transient dynamics characterizing transient dynamics.

\begin{figure*}[!htb]
\centering
\includegraphics[width=0.95\linewidth]{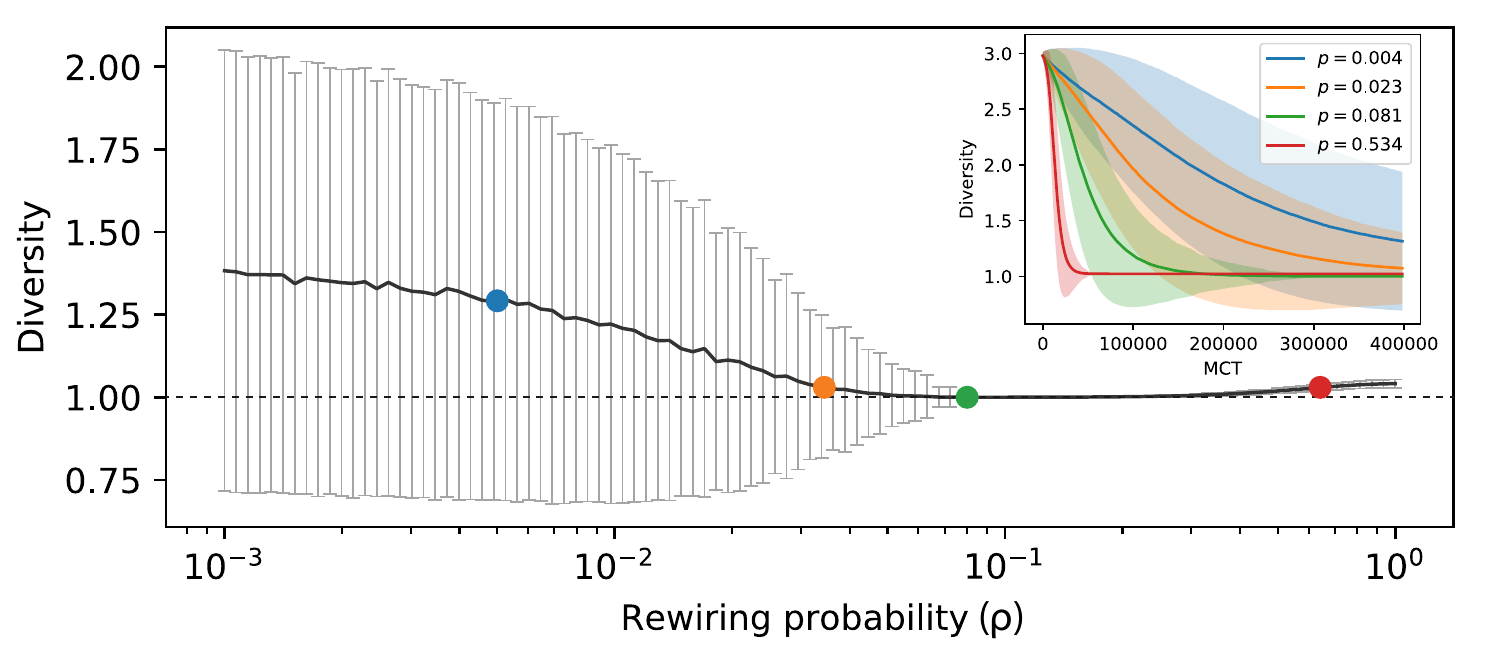}
\caption{Average opinion diversity and standard deviation as a function of rewiring probability considering 400000 Monte Carlo iterations. The inset shows how diversity changes with the number of iterations (MCT) for the four rewiring probabilities highlighted in the main plot.}
\label{fig:diversity}
\end{figure*}

Figure~\ref{fig:diversity} illustrates the results (average $\pm$ standard deviation) obtained for the above described experiments. The resulting curves can be separated into two portions: the first extending from $\rho=0$ to $\rho \approx 0.08$, the second portion unfolds from this value up to $\rho = 1$.  The first portion is characterized by steep reduction of the opinion diversity as well as of the respective standard deviation and, at $\rho \approx 0.08$, it reaches null variation and minimal average diversity (green point).  This region shows that, at least for the adopted models and configurations, the diversity reduces markedly with the loss of spatiality (as $\rho$ increases).  A possible explanation for this property is that the transient phase becomes briefer as the network looses spatiality through the rewiring procedure, as shown in the inset of Figure~\ref{fig:diversity}. Also, a large standard deviation is observed for small values of $\rho$.  The second region reveals a relatively small increase of diversity with the loss of spatiality, accompanied by a similar increase of the standard deviation of the diversity. 

In short, we can postulate that in networks predominantly spatial (i.e. with little small world characteristic), the diversity tends to increase steeply with the spatiality.  On the other hand, in small world networks such as those underlying the portion 2 of the results in Figure~\ref{fig:diversity}, the diversity increases moderately with the rewiring probability.

\section{Conclusions}
\label{sec:conclusions}

One important property of dynamics unfolding on networks concerns the diversity or heterogeneity of the respective states.  This type of characterization has direct implications on several real-world systems, such as variety of opinions, ecology, economy and transportation.  The present work argues that the spatiality of networks can promote the diversity of respective dynamics.  By performing simulations with a type of Sznajd dynamics and Watts-Strogatz networks with varying rewiring probabilities, we illustrated that the diversity is enhanced for networks that are predominantly spatial and reduces in networks characterized by more intense small world effect.
The reported study has important implications for real-world complex systems, as it indicates that, by increasing the small-world property and consequently reducing the networks spatiality, enhanced connectivity can have the effect of constraining the diversity/heterogeneity of dynamical states in several systems.  It also illustrates the fact that spatiality acts as a kind of constraint to information diffusion in dynamical complex systems.  Additional investigations need to be performed in order to identify which classes of dynamics are more affected by spatiality and addressing in more depth the transient and equilibrium properties of these systems.

\section*{Acknowledgements}
Bruno Messias thanks CAPES for financial support. Filipi N. Silva thanks FAPESP (grant no. 2015/08003-4 and 2017/09280-7) for sponsorship. Cesar H. Comin thanks FAPESP (grant no. 15/18942-8) for financial support. Luciano da F. Costa thanks CNPq (grant no. 307333/2013-2) and NAP-PRP-USP for sponsorship. This work has been supported also by FAPESP grants 11/50761-2 and 2015/22308-2 and CAPES.

\bibliographystyle{unsrt}
\bibliography{references}

\end{document}